# Valuing the quality option in agricultural commodity futures – a Monte Carlo simulation based approach

**Sanjay Mansabdar**

Department of Economics and Finance,

Birla Institute of Technology and Science, Pilani,

Hyderabad Campus,

Shamirpet Mandal, Hyderabad 500078, Telangana, India.

Tel: +91 9966883614

e mail: p20160504@hyderabad.bits-pilani.ac.in

**Hussain C. Yaganti**

Assistant Professor,

Department of Economics and Finance,

Birla Institute of Technology and Science, Pilani,

Hyderabad Campus,

Shamirpet Mandal, Hyderabad 500078, Telangana, India.

e mail: hussain@hyderabad.bits-pilani.ac.in

# Valuing the quality option in agricultural commodity futures – a Monte Carlo simulation based approach

**Abstract**

Agricultural commodity futures are often settled by delivery. Quality options that allow the futures short to deliver one of several underlying assets are commonly used in such contracts to prevent manipulation. Inclusion of these options reduces the price of the futures contract and leads to degraded contract hedging performance. Valuation of these options is a first step in assessing the impact of the quality options embedded into a futures contract. This paper demonstrates a Monte Carlo simulation based approach to estimate the value of a quality option. In order to improve simulation efficiency, the technique of antithetic variables is used. This approach can help in the assessment of the impact of embedded quality options.

## 1 Introduction

Futures contracts that are settled by delivery specify, in the contract specifications, a par asset that is the underlying deliverable asset for the futures contract at expiration. Agricultural commodity futures contracts typically allow the short at expiration to deliver one of several assets in addition to the par asset for a penalty called the discount that is applied to the settlement price calculated had the par asset been delivered. This choice of assets represents the ownership by the short of a quality option. This option is embedded into the contract definition and is traded as a package along with the futures contract.

Inclusion of these options has consequences, primarily the reduction of price of the futures contract relative to a contract that features no options (see Chance and Hemler, (1993)) and a degradation in hedging effectiveness of the futures contract (see Johnston and McConnell, (1989)). To assess the impact of these options an easily implementable method of valuing them is necessary.

These options fall under the broad class of multi asset options, where the option price depends on the value of several underlying assets. Closed form solutions for some types of multi-asset options do exist, for instance by Margrabe (1978) , Stulz (1982) and Johnson (1987). When the number of assets increases beyond two, closed form solutions for the values of such options are exceedingly complex to evaluate and some approximation algorithms, such as those suggested by Boyle and Tse (1990) have been developed. However, Monte Carlo methods are useful for multi-asset options given the complexity and accuracy constraints of other methods. Monte Carlo methods for option pricing were first used by Boyle (1977) using the technique of risk neutral valuation. Ding and Ping (2009) illustrate how Monte Carlo simulation can be applied for quanto options, one type of multi asset option.

Valuation of quality options when there are more than two underlying assets has generally followed a two-step procedure requiring the valuation of futures with and without the delivery option and taking their difference. In this paper, we illustrate the use of Monte Carlo simulation to value the quality option directly. This illustration will serve as a reference for academics and practitioners to value the quality option and research it's impact of the associated futures contract, such as the extent of price reduction and the impact on hedging effectiveness.

The remainder of this paper proceeds as follows. Section 2 briefly describes some commonly used approaches to the valuation quality options in the literature, Section 3 describes our implementation,  Section 4 contains  some numerical examples and Section 5 concludes.

## 2 Pricing the quality option

As suggested by Hrainova and Tomek (2005), the value of a delivery option is given by

$$L_{t,T} = F_{wo,t} - F_{w,t} \tag{1}$$

where $L_{t,T}$ is the value of the quality option at time $t$ for a future expiring at $T$, $F_{wo,t}$ is the value of a future that does not contain any embedded options and $F_{w,t}$ is the value of a future that contains an embedded delivery option. $F_{wo,t}$ at time $t < T$ is given by the theoretical futures pricing equation

$$F_{wo,t} = S_{1,t} e^{-r_F(T-t)} \tag{2}$$

where $S_{1,t}$ is the price of the par asset $S_1$, $r_F$ is the risk free rate and $t$ is any time prior to expiration. $F_{w,t}$ is modelled as a European call on the minimum of all the deliverable assets $S_{1,t}, S_{2,t}, \ldots, S_{n,t}$ struck at zero and with an expiration of $T$, where $S_2, S_3, \ldots, S_n$ are the additional deliverable assets in addition to the par asset.

$$F_{w,t} = \text{EC}(\min(S_{1,t}, S_{2,t}, \ldots, S_{n,t}), 0, T) \tag{3}$$

where $\text{EC}(\min(S_{1,t}, S_{2,t}, \ldots, S_{n,t}), 0, T)$ represents a European call on the minimum of the assets $S_1, S_2, \ldots, S_n$ struck at zero and expiring at time $T$. The methods of Stulz (1982), Johnson (1987), Boyle (1989) and Boyle and Tse (1990) focus on valuing $F_{w,t}$ in Equation 3 above either via a closed form solution or via some approximation. Option value can then be estimated by substitution into Equation 1 from Equations 2 and 3.

To evaluate the RHS of Equation 3, techniques seek to solve the Black Scholes equation for multi-asset options

$$\frac{\partial F_{w,t}}{\partial t} + \frac{1}{2}\sum_{i,j=1}^{n}\sigma_i\sigma_j S_i S_j \rho_{ij}\frac{\partial^2 F_{w,t}}{\partial S_i \partial S_j} + \sum_{i=1}^{n} r_F S_i \frac{\partial F_{w,t}}{\partial S_i} - r_F F_{w,t} = 0 \quad (4)$$

subject to the boundary condition

$$F_{w,T} = \max\{\min(S_{1,T}, S_{2,T}, \ldots, S_{n,T}), 0\} \quad (5)$$

where each asset $S_i$ satisfies

$$dS_i = \mu_i S_i dt + \sigma_i S_i dW_i \quad (6)$$

and

$$dS_i dS_j = \sigma_i \sigma_j S_i S_j \rho_{ij} dt \quad (7)$$

Here $\mu_i$ and $\sigma_i$ ( $i = 1,2,\ldots,n$ ) represent expected rate of return and volatility of each asset $S_i$ respectively, $dW_i$ ( $i = 1,2,\ldots n$) represents standard Brownian motion and $\rho_{ij}$ represents the correlation between assets $S_i$ and $S_j$.

Stulz (1982) developed a closed form solution that only applies to a two asset case and Gay and Manaster (1984) apply the Margrabe (1978) technique, a special case of the Stulz (1982) method. The approach of Johnson (1987) can deal with multiple assets but becomes rapidly complex as the number of underlying assets increase. Applying the technique by Boyle (1989) requires making restrictive and unrealistic assumptions pertaining to the correlations and volatilities of the deliverable assets, while the technique of Boyle and Tse (1990) uses the Clark (1961) approximation recursively, which has been shown to be acceptably accurate only under some conditions (see Greer and La Cava (1979)). Monte Carlo methods are generally used when there are multiple assets, given these challenges in using the methods described above.

**3 Monte Carlo approach**

We illustrate a Monte Carlo simulation based method to value the quality option directly, which combines the two step approach described in Section 2. Let us consider a case with three deliverable assets $S_1$, $S_2$ and $S_3$. $S_1$ is the par asset and $S_2$ and $S_3$ are the additional deliverable assets with discounts of $d_2$ and $d_3$ respectively that are applied to the settlement price received by the seller had $S_1$ been delivered, if $S_2$ or $S_3$ is delivered instead of the par asset $S_1$. On expiration, the short position holder commits to delivering the par asset $S_1$ worth $S_{1,T}$. However the embedded quality option allows the substitution of the par asset $S_1$ by $S_2$, with a penalty of $d_2$ or $S_3$ with a penalty of $d_3$. The option payoff is $S_{1,T} - S_{2,T} - d_2$ if the asset $S_2$ is delivered and $S_{1,T} - S_{3,T} - d_3$ if $S_3$ is delivered. The seller would prefer to meet his delivery obligation by maximizing his payoff by delivering the cheapest to deliver of the deliverable assets, i.e the asset that maximizes his payoff. The seller's payoff at expiration would be

$$L_{t,T} = \max(S_{1,T} - \min(S_{2,T} + d_2, S_{3,T} + d_3), 0) \qquad (8)$$

with the option only being exercised if the payoff is positive. In the event that there are totally $n - 1$ additional deliverable assets, this payoff is modified as

$$L_{t,T} = \max(S_{1,T} - \min(S_{2,T} + d_2, S_{3,T} + d_3, \dots, S_{n,T} + d_n), 0) \qquad (9)$$

We make the usual assumption of all the assets prices following correlated multivariate log normal distributions and estimate cross asset correlations and volatilities using 30 days of asset price data prior to the option valuation date. We then use a Cholesky decomposition technique to simulate 100,000 sets of asset prices at expiration $T$ assuming risk neutrality. Once the terminal asset prices are available, for each simulation we compute the terminal payoff in accordance with Equation 9. We use an antithetic variables technique to increase the efficiency of the simulation (see Boyle (1977)). Antithetic variables involves simulating, for each case, a second case that is perfectly negatively correlated to the first in order to reduce the variance. Since computational power is cheaply available in the present environment, 100,000 such simulations are run and the average value of the terminal payoffs is the estimated value of the quality option at expiration. We remark that the two subscripts assigned to the option value on the left hand side of Equation 9 remain relevant as the expiration values $S_{1,T}, S_{2,T}, \ldots, S_{n,T}$ are simulated based on the values of assets at time $t$, of $S_{1,t}, S_{2,t}, \ldots, S_{n,t}$. Additionally we observe that since the option is embedded into a futures contract, $L_{t,T}$ represents the appropriate quality option value without further discounting to time $t$, as discussed by Boyle (1989).

**4 Numerical Examples**

*A. Implementing the Boyle (1989) examples:*

We first assess the accuracy of the simulation by comparing to the results obtained using the method of Boyle (1989). As discussed in Section 2, the method focuses on computing the value of the future with the embedded quality option in Equation 3. He uses order statistics to compute the value of a future with multiple deliverable assets, all with equal prices of $40, equal volatilities of 25% (annualized) and assumes that all assets are equi-correlated (with two

values of correlation used as shown in Table 1). He uses an annual risk free rate of 10% per year and also computes the value of a future that does not contain any quality option as equal to $ 43.11. He also assumes zero discounts, equi-correlated multivariate log normal distribution of asset prices and reports the ratio of the futures price containing the quality option to the futures price without the option. From his results, we calculate the value of the quality option using Equation 1 and compare this to the results of the Monte Carlo simulation described above applied with the asset prices, correlations and volatilities use by Boyle(1989) in Table 1. We see that the relative error is generally within 0.25% of the values of Boyle (1989).

*Table 1:Comparing the results of Boyle (1989) to those of Monte Carlo simulation.*

| Total Number of Deliverable Assets (n) | Correlation Coefficient = 0.95 | | | Correlation Coefficient = 0.995 | | |
|---|---|---|---|---|---|---|
| | Quality Option Value* in $ (Boyle) | Quality Option Value* in $ (MC) | Relative Error (percent) | Quality Option Value* in $ (Boyle) | Quality Option Value* in $ (MC) | Relative Error (percent) |
| 2 | 1.117 | 1.118 | 0.09 | 0.371 | 0.372 | 0.27 |
| 3 | 1.750 | 1.749 | -0.06 | 0.556 | 0.557 | 0.18 |
| 4 | 2.121 | 2.124 | 0.14 | 0.677 | 0.676 | -0.15 |
| 5 | 2.389 | 2.386 | -0.13 | 0.763 | 0.763 | 0.00 |
| 10 | 3.126 | 3.121 | -0.19 | 1.009 | 1.008 | -0.10 |
| 20 | 3.760 | 3.763 | 0.08 | 1.220 | 1.220 | 0.00 |
| 30 | 4.096 | 4.090 | -0.14 | 1.332 | 1.331 | -0.08 |
| 40 | 4.319 | 4.317 | -0.05 | 1.406 | 1.407 | 0.07 |
| 50 | 4.484 | 4.482 | -0.04 | 1.462 | 1.463 | 0.07 |
| Quality options are calculated using both methods for a future with equi-correlated deliverable assets, with initial values of $40, standard deviations of 25% per year and a risk free rate of 10% per year with the indicated number of deliverable assets and correlations assumed. | | | | | | |

*B. Valuing options for Chana August 2014 contract on the National Commodities and Derivatives Exchange of India (NCDEX):*

We illustrate the simulation for the agricultural contract traded on the NCDEX in India. We choose this market and contract for the illustration because the relevant time series of cash and futures market prices are readily available for all the deliverable assets. Chana contracts were and remain among the more liquid contracts traded on the exchange. The contract expiry date is 20$^{th}$ August 2014. The contract had three delivery centres, Delhi, Bikaner and Indore, with the cash price at Delhi representing the par asset and the cash prices at Bikaner and Indore representing the additional delivery asset. We value the quality option on specific dates in Table 2 during the time when the futures contract itself is liquid, which are the two calendar months prior to the expiration month, June 2014 and July 2014 in this case. The risk free rate is assumed as 7.5% per year. The following data are available from the NCDEX website

1. Daily Cash prices of Chana in the three delivery centres for May, June and July 2014
2. Daily Futures prices for the August 2014 contract during June and July 2014
3. Discounts applicable at Bikaner and Indore for the July contract (at Rs 70 and Rs 19 respectively).

On each business day (the valuation date) during June and July 2014, the previous 30 day prices are used to estimate correlations and volatilities of asset returns. Closing prices in each of the deliverable cash markets on the day prior to the option valuation date are used as initial asset prices. Using the simulation procedure outlined in Section 2, quality option values for that valuation date. The quality option values computed are reported on a standalone basis and as a ratio of the futures price on the valuation date. Log normal multivariate distribution of cash market prices is assumed and options are valued assuming no delivery before expiration. Table 2 provides a summary of the results of the valuation using Monte Carlo simulation. These are perhaps the first available estimates of quality option value for Indian agricultural commodity futures.

*Table 2: Illustration of Quality Option Value and Quality Option Value as a ratio of futures price calculated using Monte Carlo simulation for the August 2014 Chana contract traded on the NCDEX for different trading days.*

| Valuation Date | Futures Price (Rs) | Delhi Cash Price* (Rs) | Bikaner Cash Price* (Rs) | Indore Cash Price* (Rs) | Quality Option Value (Rs) | Quality Option as a percentage of Futures price |
|---|---|---|---|---|---|---|
| 2-Jun-2014 | 2913 | 2813.25 | 2810.00 | 2697.05 | 141.80 | 4.87 |
| 3-Jun-2014 | 2957 | 2800.00 | 2810.00 | 2678.15 | 143.12 | 4.84 |
| 3-Jun-2014 | 2985 | 2843.75 | 2845.00 | 2731.50 | 127.12 | 4.26 |
| 4-Jun-2014 | 2960 | 2838.45 | 2830.00 | 2737.40 | 118.21 | 3.99 |
| 5-Jun-2014 | 2915 | 2800.00 | 2790.00 | 2700.65 | 112.93 | 3.87 |
| *Cash Prices shown are actual cash prices plus applicable discount. | | | | | | |

## 5 Conclusions

This paper illustrates the use of Monte Carlo simulation to estimating the value of the quality option embedded in futures contracts directly, when other methods are difficult to implement, require unrealistic assumptions or are not accurate enough. The approach is intuitive, easy to implement and sufficiently accurate in a world where computational costs are low. As such this paper should serve as a reference for contract designers, practitioners and academics who need to estimate the value of the quality option.